\documentclass[twocolumn,showpacs,preprintnumbers]{revtex4}
\usepackage{amssymb}
\usepackage{graphicx}
\usepackage{dcolumn}
\usepackage{bm}

\begin{document}

\preprint{}
\title{An optical phase-locking with large and tunable frequency difference based on vertical-cavity surface-emitting laser
}
\author{Wenlan Chen}
\author{Xianghui Qi}
\author{Lin Yi}
\author{Ke Deng}
\author{Xuzong Chen}
\thanks{Corresponding author: xuzongchen@pku.edu.cn}

\affiliation{Institute of Quantum Electronics, \\ School of
Electronics Engineering and Computer Science,
\\Peking University, Beijing, 100871, China}
\date{\today }

\begin{abstract}
We present a novel technique to phase-lock two lasers with
controllable frequency difference. In our setup, one sideband of a
current modulated Vertical-Cavity Surface-Emitting Laser (VCSEL) is
phase locked to the master laser by injection seeding, while another
sideband of the VCSEL is used to phase lock the slave laser. The
slave laser is therefore locked in phase with the master laser, with
a frequency difference tunable up to about 35 GHz. The sideband
suppression rate of the slave laser is more than 30dB at 30 $\mu$W
seed power. The heterodyne spectrum between master and slave has a
linewidth of less than 1 Hz. A narrow linewidth spectrum of coherent
population trapping in rubidium is achieved using such beams.
\end{abstract}


\maketitle

\noindent There have been a number of reports related to atomic
coherence effects, such as atomic clocks based on coherent
population trapping \cite{contrastCPT, CPTclock},
electromagnetically induced transparency \cite{EIT}, and Raman
cooling \cite{Cooling}. In these experiments, the requirement of the
relative frequency stability of two lasers exceeds that of absolute
frequency by far. A tunable frequency difference of several GHz is
necessary and can be obtained by stabilizing one laser to the other
in a heterodyne optical phase-locked loop. In this approach, the
beat note between the two lasers is detected and phase-locked to a
local oscillator by an electronic feedback loop. The requirements to
lock the laser are a phase locked loop operating at the offset
frequency, typically several GHz for alkali atoms in the
ground-state, and a wide-bandwidth feedback \cite{ELock}.

Alternatively, in an all-optical loop such beams can be generated in
a simpler way. After modulation using an electro-optic modulator
(EOM) or an acousto-optical modulator (AOM) or an Fabry-Perot (FP)
laser diode, one sideband of the laser source is used to lock a
slave laser by injection seeding \cite{Filter, Filter2,
DecreasePower, DiodeMod}. This way, the slave laser is optically
phase locked to the master laser with a frequency difference
determined by the modulation frequency. But using these methods,
because of the low efficiency of such modulators, the locked slave
laser would have an unwanted oscillating whose frequency is the same
as the carrier of the master laser. In order to remove this unwanted
mode, there was a suggestion to use a locked filtering cavity or to
decrease the seed power \cite{Filter, Filter2, DecreasePower}.
However, the filtering cavity makes the laser system complicated
while the decreased power reduces the reliability and the locking
range of the injection locking.

\begin{figure}[t]
\centerline{\includegraphics[width=8.3cm]{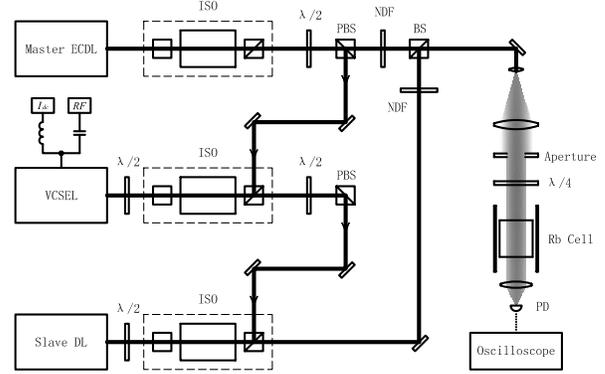}}
\caption{Experimental setup for producing phase-coherent laser beams
with large frequency difference and obtaining CPT spectrum of the
$5S_{1/2}-5P_{3/2}$ transition of $^{87}$Rb using such beams. ISO:
isolator; $\lambda/2$: half-wave plate; PBS: polarizing beam
splitter; BS: beam splitter; NDF: neutral density filter;
$\lambda/4$: quarter-wave plate; PD: PIN photo diode.}\label{setup}
\end{figure}

In this paper, we demonstrated a convenient approach to produce
optical phase locked laser beams of high frequency purity, large
frequency difference and wide dynamic locking range. Instead of an
AOM, EOM or FP laser diode, we used a vertical-cavity
surface-emitting laser (VCSEL) diode as modulator in the injection
locking process. The current driving a commercial VCSEL can be
modulated at frequencies up to several GHz. In contrast to the
limited modulation efficiency of EOMs, AOMs or FP diodes, the
sideband intensity generated by a VCSEL could be even higher than
the carrier intensity. As a result of this high efficiency
modulation, carrier of master laser is depressed in the slave laser
even at high seed power. Since high seed power is possible, the
frequency difference tuning can be accomplished in a wide,
continuous range and at high speed, which also benefit from mode-hop
free feature of VCSELs. Further more, optical phase locking to the
higher-order sidebands is also available, which further extends the
frequency difference range: using the upper third and lower second
harmonics, it is able to obtain a frequency difference of 35 GHz at
7 GHz current modulation.

The experimental setup is shown schematically in Fig. \ref{setup}. A
VCSEL plays the essential role in phase-locking a slave laser to the
master laser. This VCSEL (PS78-F1P0N) is phase-modulated by a radio
frequency current modulation in order to produce sidebands. One of
these sidebands is injection-locked to the master
laser\cite{DiodeMod}, which is an external-cavity diode laser of 300
kHz linewidth (Vortex 6013). Due to the low VCSEL output power of
about 2 mW, this locking is working at a low seed power of about 5
$\mu$W. By adjusting the VCSEL current, any sideband could be locked
to the master laser (except very weak ones). In the same way by
adjusting the slave laser current, the slave laser (FP laser diode)
frequency is injection-locked to another sideband or to the carrier
of the VCSEL at a seed power of several tens of $\mu$W. As a result,
carrier of the slave laser is phase-locked to the master laser, with
30 mw output power in our case, and with the frequency difference
amounts to several times of the modulation frequency.

\begin{figure}[t]
\centerline{\includegraphics[width=7cm]{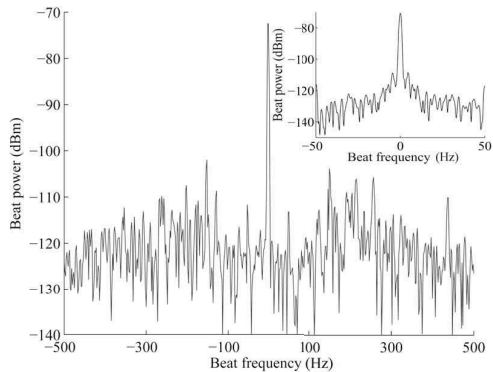}}
\caption{Beat-note spectrum at 6.8GHz between master laser and
locked slave laser. Span: 1 kHz; RBW: 3 Hz; VBW: 3 Hz; SWP: 808 ms.
The inset shows a smaller span. Span: 100 Hz; RBW: 1 Hz; VBW: 1 Hz;
SWP: 2.98 s. }\label{Beat}
\end{figure}

To measure the phase-coherence between the master laser and the
slave laser, we detected the beat-note spectrum of them with a fast
PIN photo diode (Hamamatsu G4176), while the master laser was 110
MHz shifted by an AOM (AA.MT.110/B50/A1.5-IR). The relative
linewidth of the two lasers was observed using a RF spectrum
analyzer (HP8563E). When phase-locked, the relative linewidth was
measured to be 1 Hz as shown in Fig. \ref{Beat}, which was at the
resolution limit of the spectrum analyzer. This linewidth is
estimated much smaller than the width of coherent population
trapping resonance of most atoms. At the same time, the absolute
linewidths of the slave laser was decreased to about 300 kHz, the
same as the master laser's, which depresses the large initial
linewidth of VCSELs.

\begin{figure}[t]
\centerline{\includegraphics[width=7cm]{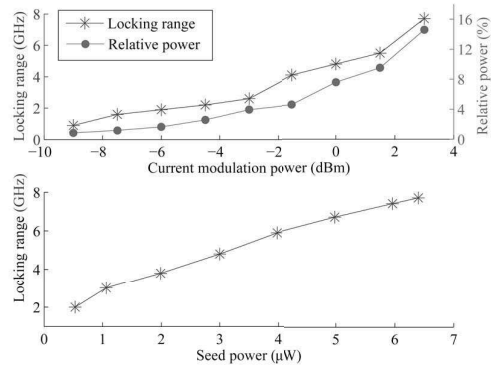}}
\caption{Locking range of VCSEL as a function of current modulation
power and injection seed power. (a) Locking range increases with the
enhancement of modulation power (stars) and thus fractional laser
power in the upper first-order sideband (solid circles) when the
seed power is fixed at 6.4 $\mu$W. (b) Locking range as a function
of seed power when modulation power is 3 dBm.}\label{LockRange}
\end{figure}

Due to VCSELs' architecture, they are inherently single longitudinal
mode and can be continuously tuned for a wide range by raising the
drive current at a fixed temperature \cite{ModeHopFree}. This
feature of far wavelength tuning without mode hops brings
convenience and large locking range in locking VCSEL to the master
laser. Free from bother of mode-hops, we measured the locking range
as a function of the seed power and fractional laser power in the
upper first-order sideband of VCSEL (Fig. \ref{LockRange}) when this
sideband was locked to the master laser. When seed power or relative
sideband power increases, the locking range increases accordingly.
With 6.4 $\mu$W seed power and 14.6\% relative sideband power, the
locking range can be as wide as 7.7GHz. It increases much slower
after 7.7GHz, cause at the lower frequency boundary, the master
laser would lock the carrier of VCSEL instead of the sideband, which
is reasonable considering the modulation frequency of 6.8GHz.

To prove the spectral purity of the slave laser, we observed it with
a Fabry-Perot interferometer having a free spectral range of 4 GHz.
When the VCSEL was modulated at 6.8 GHz with a modulation power of 3
dBm, we locked the first order upper sideband of the VCSEL to the
master laser (5 $\mu$W seed power), and injected 30 $\mu$W laser
power to lock the slave laser (5 mW output power) to the carrier of
VCSEL. Under these conditions, our setup could work steadily without
off locking for several days and the sideband intensity of the slave
laser was below $-$30 dB of the carrier intensity. Therefore,
negative effects such as inducing an AC stark shift in a CPT clock
or selecting atoms with a wrong velocity in laser cooling can be
avoided.

We further tested the tuning speed and tuning range of the frequency
difference. We made sure that the coherence between master and slave
laser is good by measuring the beat-note specturm intensity. At a
tuning rate of 32 MHz/$\mu$s, the highest rate offered by our signal
generator (Agilent E8257D), this coherence remained good. The
continuous tuning range of frequency difference between master and
slave was about 2 GHz at 3 $\mu$W seed power for the VCSEL and 30
$\mu$W for the slave laser, while all the temperature and current
settings were fixed.

Due to the small active layer volume and high photon density, VCSELs
can be modulated at much higher frequency than edge-emitting diode
lasers \cite{HighMod}. We take advantage of this efficient
modulation: at a modulation frequency as high as 7 GHz for a
commercial VCSEL of 2.5 Gbps, the intensity of the first order
sidebands could be even higher than that of the carrier. Since
VCSELs have very short cavity of several micrometers in optical
length and relevant large laser linewidth of several hundred Mega
Hertz, they could tolerate higher order sidebands of high frequency
modulation to oscillate in them. Thus more than 8 sidebands could
emerged in the VCSEL frequency with proper current modulation power.
We have chosen the third upper sideband to be locked to the master
laser, and the second lower sideband to lock the slave laser, and
obtained a frequency difference of 35 GHz between the two lasers.

\begin{figure}[t]
\centerline{\includegraphics[width=7cm]{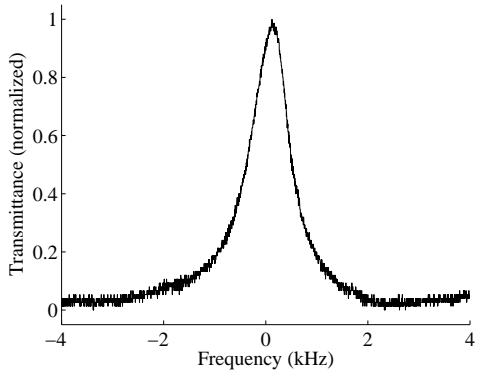}}
\caption{Lambda-type CPT spectrum of the $5S_{1/2}--5P_{3/2}$
transition of $^{87}$Rb obtained two phase-locked lasers. The
linewidth of the CPT spectrum is 800Hz.}\label{800HzCPT}
\end{figure}

The system described in this article was used to observe the CPT
resonance of the $5S_{1/2}-5P_{3/2}$ transition of $^{87}$Rb. Using
a saturated absorption technique, we fixed the frequency of the
master laser at the $F=1 \rightarrow F'=1$ transition. While the
VCSEL was modulated at near 6.8 GHz (the split of the ground state
$F=1$ and $F=2$ of $^{87}$Rb), its first order upper sideband was
locked to the master, and its carrier was used to lock the slave
laser. Thus slave laser was phase-locked to the master laser with a
frequency difference controlled by VCSEL's current modulation
frequency. The master and slave laser were combined to pass through
a rubidium vapor cell (4 cm length, argon-neon buffer gas at room
temperature). Then CPT spectrum can be detected by a photodiode as
the frequency difference was scanned. The intensity of the two beams
was 25 $\mu$W/cm$^2$ each. In Fig. \ref{800HzCPT}, the CPT spectrum
shows the linewidth of 800 Hz, which is mainly limited by collision
broadening and saturated broadening. Comparing with obtaining the
CPT spectrum simply using a modulated VCSEL, this setup can control
the intensity of the two frequency components separately, and easily
avoided the light shift arising from the undesired off-resonant
fields \cite{lightshift}.

To summarize, we have demonstrated a simple and reliable scheme to
obtain two separate laser beams with large and tunable frequency
difference and high coherence. Using a VCSEL as the modulator and by
twice injection locking, we experimentally obtained a frequency
difference up to 35 GHz, and a beat-note between the two beams of
less than 1 Hz linewidth. The sideband suppression rate of the
locked laser was more than 30 dB. The frequency difference could be
tuned continuously over 2GHz and at high rate. Since far wavelength
tuning without mode hops is possible for VCSELs, this system shows
high reliability and more convenience. CPT resonance of rubidium was
observed as the frequency difference between master and slave laser
was scanned.

In optical frequency metrology, we can use this scheme to improve
the performance of direct frequency comb measurements of absolute
optical frequencies \cite{FSMeas}. Using a locked femtosecond comb
as master laser, the VCSEL and the following slave laser can filter
out a single comb mode and largely increase its intensity while
remaining coherent with such mode
\cite{CombInject1,CombInject2,CombInject3}. At the same time, we can
control the shift of the slave laser's frequency relative to the
comb mode frequency. Considering VCSELs' mode-hop free behavior and
low seed power demand, this becomes a convenient supplement to scan
and measure spectra frequencies while keeping femtosecond comb
fixed, and benefits high accuracy absolute frequency measurement of
large amount of transitions in atoms and molecules.

This work was supported by the National Natural Science Foundation
of China (Grant No. 60490280, 69789801, 10074003), the state Key
Development Program for Basic Research of China (Grant No.
2001CB309308) and the National Hi-Tech ICF program.

\end{document}